\documentclass[preprints,article,accept,moreauthors,pdftex]{Definitions/mdpi} 
\firstpage{1} 
\pubvolume{1}
\issuenum{1}
\articlenumber{0}
\pubyear{2022}
\copyrightyear{2022}
\datereceived{} 
\dateaccepted{} 
\datepublished{} 
\hreflink{https://doi.org/} 

\usepackage{amsmath}
\usepackage{comment}
\usepackage{empheq}
\usepackage{MnSymbol}

\DeclareMathOperator{\arsinh}{arsinh}
\DeclareMathOperator{\arcosh}{arcosh}
\DeclareMathOperator{\Tr}{Tr}

\newcommand{\Sk}{ S_{\kappa}}

\newcommand{\ku}{ u_{\kappa} }
\newcommand{\kexp}{ \exp_{ \kappa } }
\newcommand{\kln}{ \ln_{\kappa} }
\newcommand{\ave}[1]{\left\langle #1 \right\rangle}

\newcommand{\kExp}{ {\rm Exp}_{ \kappa } }

\newcommand{\ksum}{ \overset{\kappa}{\oplus} }

\newcommand{\Ev}[2]{{\mathbb E}_{{#1}} \left[ #2 \right]}

\Title{ 
On the Kaniadakis distributions applied in statistical physics and natural sciences}

\TitleCitation{On the Kaniadakis distributions}

\Author{Tatsuaki Wada $^{\dagger}$\orcidA{}, and Antonio. M. Scarfone $^{\ddagger}$\orcidB{}}

\AuthorNames{Tatsuaki Wada, and Antonio. M. Scarfone}

\AuthorCitation{Wada, T.; Scarfone, A.M.}

\address{%
$^{\dagger}$ \quad Region of Electrical and Electronic Systems Engineering, Ibaraki University, 
 Nakanarusawa-cho, Hitachi-shi, Ibaraki, 316-8511, Japan; tatsuaki.wada.to@vc.ibaraki.ac.jp\\
$^{\ddagger}$ \quad Istituto dei Sistemi Complessi, Consiglio Nazionale delle Ricerche (ISC-CNR), 
c/o Politecnico di Torino, Corso Duca degli Abruzzi, 24, 10129, Torino, Italy
; antonio.scarfone@to.it
}

\corres{Correspondence: tatsuaki.wada.to@vc.ibaraki.ac.jp}

\abstract{
Constitutive relations are fundamental and essential to characterize physical systems. By utilizing the $\kappa$-deformed functions, some constitutive relations are generalized. We here show some applications of the Kaniadakis distributions based on the inverse hyperbolic sine function to some topics belonging to the realm of statistical physics and natural science.}

\keyword{$\kappa$-deformed functions; constitutive relations; Gompertz rule;  Lotka-Volterra equations; contact density  dynamics
} 

\begin{document}



\section{Introduction}


The $\kappa$-exponential function \cite{KS02,K02,K05} is defined by

\begin{align}
  \kexp(x) := \left( \kappa x + \sqrt{1 + \kappa^2 x^2} \right)^{\frac{1}{\kappa}} = \exp \left[ \frac{1}{\kappa} \arsinh \left( \kappa x \right) \right],
 \label{k-exp}
\end{align}
for a real deformation parameter $\kappa$. The inverse function, i.e., the $\kappa$-deformed logarithmic function, is defined by

\begin{align}
  \kln x := \frac{x^{\kappa} - x^{-\kappa}}{2 \kappa} =   \frac{1}{\kappa} \sinh \left[  \kappa \ln x \right].
  \label{k-ln}
\end{align}
Both $\kappa$-deformed functions are important ingredients of the generalized statistical physics based on $\kappa$-entropy 
\cite{KS02,K02,K05}. 
It influences a wide range of scientific fields, and based on the $\kappa$-deformed functions (Appendix \ref{Appendix A}), several basic fields have been developed over two decades.
Kaniadakis \cite{K13} provided the theoretical foundations and mathematical formalism generated by the $\kappa$-deformed functions, and some references including many fields of applications.
Recently, the usefulness of the $\kappa$-statistics was demonstrated for the analysis \cite{Kepi} of epidemics and pandemics.


Constitutive relations are fundamental and essential to characterize physical systems. They are combined with the other equations of the physical laws in order to solve physical problems. Some well-known examples of linear constitutive relations are: Hooke's law $F = k_s x$ for the tensile or compressive force $F$ of a spring with a spring constant $k_s$ against the change in its length $x$; Ohm's law $V = R I $ for the voltage $V$ of a electrical conductor with resistance $R$ under an electric current $I$, and so on. However, as a real spring deviates from Hooke's law, we know that   
 any linear constitutive relation describes an idealized situation, and it is merely a linearized- and/or approximated- relation to describe some real physical properties. Hence, in general, non-linearity plays a crucial role to describe more realistic physical systems.

The $\kappa$-exponential function \eqref{k-exp}  can be regarded as a useful tool (or device) to make such non-linear constitutive relations for a better description of real physical systems.
For example, consider the following $\kappa$-deformation of Hooke's law:

\begin{align}
       F_{\kappa} := k_s \ln\left[ \kexp(x) \right] = \frac{k_s}{\kappa} \ln \left( \kappa x + \sqrt{1 + \kappa^2 x^2} \right),
\end{align}
which reduces to the original Hooke's law $F = k_s x$ in the limit of $\kappa \to 0$.
For any linear constitutive relation, we can apply this type of the $\kappa$-deformation. For example, Ohm's law can be cast into the following form:
$ V = R I =  R  \ln \left[  \exp(I) \right]$.
By changing the exponential function with the $\kappa$-exponential function, we obtain the $\kappa$-deformed version of Ohm's law:
$ V_{\kappa} =  R  \ln \left[  \kexp(I) \right]$.  In this research, we focus on this type of the $\kappa$-deformation of a physical quantity (say $A$), i.e,

\begin{align}
      A \quad \Rightarrow \quad \ln \left[ \kexp(A) \right] = \frac{1}{\kappa} \arsinh \left( \kappa  A \right).
\end{align}
Throughout in this paper, we call this $\kappa$-deformation as \textit{the $\arsinh$-type deformation} of a physical quantity $A$.
The other type of the $\kappa$-deformation can be

\begin{align}
      A \quad \Rightarrow \quad \kln \left[ \exp(A) \right] = \frac{1}{\kappa} \sinh \left( \kappa  A \right),
\end{align}
which is called here \textit{the $\sinh$-type deformation}.
In Ref. \cite{W04}, the thermodynamic stability of the $\kappa$-generalization $S_{\kappa}^{\rm B}$ of Boltzmann entropy $S^{\rm B}$ is studied.
The $\kappa$-generalization $S_{\kappa}^{\rm B}$ is rewritten in the form:

\begin{align}
  S_{\kappa}^{\rm B} := k_{\rm B} \kln W =  k_{\rm B} \kln [ \exp( \ln W) ] =  k_{\rm B} \kln[ \exp( S^{\rm B}) ],
\end{align}
which can be regarded as the $\sinh$-type deformation of Boltzmann entropy $S^{\rm B}$.
Recently, in cosmology, Lymperis et al. \cite{LBS21}  modified Bekenstein-Hawking entryopy $S^{\rm BH}$ as follows:

\begin{align}
  S^{\rm BH}_{\kappa}  =\frac{1}{\kappa} \sinh \left( \kappa  S^{\rm BH} \right),
\end{align}
which is obviously the $\sinh$-type deformation of $S^{\rm BH}$.

In this paper we consider the $\arsinh$-type deformations against some constitutive relations in the field of statistical physics and natural sciences.
In our previous work \cite{WSM20} we studied a thermal particle under a velocity-dependent potential which
can be regarded as a deformation of Rayleigh’s dissipation function  \cite{Rayleigh} and showed that the probability distribution function (pdf) for the stationary-state of this thermal particle is a $\kappa$-deformed Gaussian pdf.
It was considered the canonical pdf $\rho(v)$, in the velocity space, of a thermal particle with unit mass ($m=1$)
in the $\kappa$-deformed confining potential $U_{\kappa \beta}(v)$:

\begin{align}
U_{\kappa \beta}(v)  &:= \frac{1}{\kappa \beta} \arsinh \left(\kappa  \beta \frac{v^2}{2} \right),
\end{align}
where $\beta := 1 / k_{\rm B} T$ is a coldness (or inverse temperature).
This $\kappa$-deformed potential $U_{\kappa \beta}(v)$ is rewritten, in the momentum-space, as

\begin{align}
 U_{\kappa \beta}(p) = \frac{1}{\kappa \beta} \arsinh \left(\kappa  \beta \frac{p^2}{2} \right) = \frac{1}{\beta} \ln \left[\kexp \left( \beta \frac{p^2}{2} \right) \right] ,
\end{align}
which is the $\arsinh$ type deformation of the quantity $\beta p^2 / 2$ (the ratio of the kinetic energy to the mean thermal energy $k_{\rm B} T = 1/\beta$).
In other words, we consider the following $\kappa$-deformation $ Q_{\kappa}(U)$ of the Boltzmann factor $\exp(-\beta U)$ for an equilibrium state with the energy $U$:

 \begin{align}
   Q_{\kappa}(U) := \kexp( -\beta U)  = \exp \left[ \frac{1}{\kappa} \arsinh \left( -\kappa \beta U \right) \right].
   \label{kBfactor}
 \end{align}
One may wonder why the inverse hyperbolic sine function ($\arsinh$) plays a role. In many different fields of sciences, there is no doubt that the exponential and logarithmic functions are important and fundamental. Since the inverse hyperbolic sine function and logarithmic function are mutually related as:

\begin{align}
 \arsinh  x =  \ln \left[ x + \sqrt{1+x^2} \right],  \quad  \ln x = \arsinh \left[  \frac{1}{2} \left( x -\frac{1}{x} \right) \right],
\end{align}
for a positive real $x$,
we think both functions are important.
By using the second relation, for any real parameter $\kappa \ne 0$, we have

\begin{align}
  \ln x =  \frac{1}{\kappa} \ln x^{\kappa} = \frac{1}{\kappa} \arsinh \left[  \frac{1}{2} \left( x^{\kappa} -x^{-\kappa} \right) \right]
  =  \frac{1}{\kappa} \arsinh \left[  \kappa \kln x \right].
  \label{lnz}
  \end{align}
Note that this relation corresponds to the $\arsinh$-type deformation of $\kln x$ and is equivalent to definition \eqref{k-ln} of the $\kappa$-deformed logarithmic function that can be regarded as the $\sinh$-type of $\kappa$-deformation of $\ln x$.
Kaniadakis already discussed this issue in section II of Ref. \cite{K02} from the viewpoint of the deformed algebra.

On the other hand, Pistone \cite{P09} was the first one to study the $\kappa$-exponential model in the field of information geometry \cite{Amari},
and later through our research activities \cite{WS15,WSM20,WSM21}, we realized that there exist some relations among statistical physics, thermodynamics, mathematical biology, and information geometry. Harper \cite{Harper09,Harper11} pointed out that the replicator equation (RE) \cite{Sigmund87} in mathematical biology or in an evolutional game theory \cite{HS98} is related
with information geometry  and a general form of the Lotka-Volterra  (gLV) equation as briefly explained in Appendix \ref{Appendix B}.
The gLV equations \cite{HF97,Harper09,Harper11,Baez}:

\begin{align}
   \frac{d y_i}{dt} = y_i \; f_i(\boldsymbol{y}),
   \label{gLVeq}
\end{align}
are used to model the competition dynamics of the populations $y_1, y_2, \dots, y_n$ of $n$ biological species.
Gompertz function \cite{Gompertz} is a type of mathematical model for a time evolution. Historically he  studied human mortality and proposed his law of human mortality in which he assumed that person's resistance to death decreases as his years increase.
His law is now called \textit{Gompertz rule} (or law) and we would like to point out the relation of his function and his rule to some important quantities
concerning statistical physics.

The rest of the paper is organized as follows. In section 2, we briefly explain Gompertz function,  and the gLV equations, which are important in mathematical biology (or evolutional game theory). Their relations to thermal physics are pointed out.
Section 3 considers the thermal density operator, which is characterized by the so-called Bloch equation \cite{Bloch,Kirkwood} for thermal states, and we shall show that the Bloch equation can be regarded as a Gompertz rule after the parameter transformation $\beta$ to $t=-\ln \beta$.
In section 4, we discuss the $\arsinh$-type deformation from the viewpoint of the $\kappa$-addition.
In section 5, we study the numerical simulations of the thermostat algorithm for the Hamiltonian with the $\kappa$-deformed kinetic energy, which can be regarded as
the $\arsinh$ type of the $\kappa$-deformation of the ratio $\beta p^2/2$ as shown in \eqref{kBfactor}.
The final section is devoted to our conclusions.

\section{Gompertz functions and Gompertz rule}
Here we would like to point out that there exist relations between an evolutional game dynamics and thermal physics. 
In evolutional game theory \cite{HS98}, an evolutional game dynamics is described by a RE.  The gLV equations are related to REs as shown in Appendix \ref{Appendix B}.
On the other hand, Gompertz function is a mathematical model describing an evolutional curve.
Gompertz function (or Gompertz curve) \cite{Gompertz} is a type of mathematical model for a time series.
Gompertz function $f_{\rm G}(t)$ is a sigmoid function and is given by

\begin{align}
  f_{\rm G}(t) := K \exp\big[ C \, \exp(-t) \big],
  \label{Gfunc}
\end{align}
where $C$ and $K$ are positive constants. A distinctive feature of Gompertz function is its double exponential $t$-dependency.
His function is nowadays used in many different areas to model a time evolution of the populations where growth is slowest at the start and end of a period. For example, Ref. \cite{Gmodel} applied Gompertz model to describe the growth dynamics of COVID-19 pandemic. 
Gompertz  \cite{Gompertz} studied human mortality for working out a series of mortality tables, and this suggested to him his law of human mortality in which he assumed that the person's resistance to death decreases as the age increases. 
The rule of his model is called \textit{Gompertz rule} which states that

\begin{align}
   \frac{d}{dt} f_{\rm G}(t) = -f_{\rm G}(t) \ln \frac{f_{\rm G}(t)}{K}.
   \label{Grule}
\end{align}
The solution of the Gompertz rule is the Gompertz function \eqref{Gfunc}, if we set $K = \lim_{t \to \infty} f_{\rm G}(t)$ and $C = \ln (f_{\rm G}(0)  / K )$.

If we choose $f_i(\boldsymbol{y}(t)) = -\ln y_i(t)$ and assuming $\lim_{t \to \infty} y_i(t) = 1$, the gLV equation \eqref{gLVeq} becomes

\begin{align}
   \frac{d y_i(t)}{dt} = -y_i(t) \; \ln y_i(t),
\end{align}
which can be regarded as the Gompertz rule \eqref{Grule} with $K=1$ for each $y_i(t)$.
Consequently, its solution $y_i(t)$ is the Gompertz function:

\begin{align}
  y_i(t) = \exp\big[ \ln y_i(0) \, \exp(-t) \big].
  \label{y_i}
\end{align}
Now, by changing the parameter $t$ to $\beta = \exp(-t)$, we have $d\beta = - \beta dt$ so that the limit $t \to 0$ corresponds to $\beta \to 1$, and each constant $E_i$ are introduced as

\begin{align}
   -E_i = \lim_{t \to 0} \ln y_i(t) = \lim_{\beta \to 1} \ln y_i(\beta),
\end{align}
where $y_i(\beta)$ is the shorthand notation of $y_i(t(\beta))$ with $t(\beta)= -\ln \beta$.
Then, the solution $y_i(\beta)$ in \eqref{y_i} can be expressed as a quantity
very familiar in statistical physics:

\begin{align}
   y_i(\beta) = \exp( -\beta E_i),
\end{align}
that is Boltzmann factor.
The corresponding Gompertz rule \eqref{Grule} for $y_i(\beta)$ is equivalent to

\begin{align}
   \frac{d}{d \beta} y_i(\beta) = -E_i \, y_i(\beta).
\end{align}

Having described the relation between Gompertz rule and Boltzmann factor $\exp( -\beta E_i)$ in statistical physics,
in the next section we shall discuss a $\kappa$-deformation of Bloch equation for thermal states.

\section{Bloch equation for thermal states}
For a given Hamiltonian $\hat{H}$ and the corresponding eigenvalues $E_i$ and eigenstate  $\vert \psi_i \rangle$, which are related in

\begin{align}
  \hat{H}  \vert \psi_i \rangle =  E_i \vert \psi_i \rangle,
  \label{eigen}
\end{align}
and assuming the completeness relation $\sum_i \vert \psi_i \rangle \langle \psi_i \vert = \hat{1}$,
the density operator $ \hat{\rho}( \beta) $ for a canonical ensemble is constructed as

\begin{align}
  \hat{\rho}( \beta) :=  \sum_i  \exp( - \beta  E_i) \vert \psi_i \rangle \langle \psi_i \vert =  \exp( - \beta  \hat{H}).
\end{align}
In order to determine the canonical density matrix, we have to solve the eigenvalue equations \eqref{eigen} and to sum over all the states.
This needs heavy calculations in general. 
Note that $ \hat{\rho}( \beta) $ is un-normalized and its trace is $ \Tr \hat{\rho}( \beta) =   Z( \beta)$, which is the partition function.

Bloch equation \cite{Bloch,Kirkwood} for thermal states is known as

\begin{align}
  -\frac{\partial}{\partial \beta} \hat{ \rho}( \beta) = \hat{H} \, \hat{\rho}(\beta),
  \label{Bloch eq}
\end{align}
which can be regarded as the diffusion equation in imaginary time $\beta$, and it has a similar form as Schr\"odinger equation and diffusion equation. 
Bloch equation  \eqref{Bloch eq}  offers an alternative route to determine the density operator $\hat{\rho}(\beta)$. The initial ($\beta=0$) condition is provided if we know the eigenstates in the high-temperature limit. 

Now, by multiplying $\beta$ to both sides of \eqref{Bloch eq}, we have

\begin{align}
  -\beta \frac{\partial}{\partial \beta}  \hat{\rho}(\beta) = \beta \hat{H} \, \hat{\rho}(\beta) = -\ln[ \hat{\rho}(\beta) ] \, \hat{ \rho}(\beta).
\end{align}
Changing the parameter $\beta $ to $t = - \ln \beta$, it follows

\begin{align}
  \frac{d }{dt} \hat{\rho}(t) = -  \beta \frac{d}{d \beta} \hat{\rho}(\beta) = -\ln[ \hat{\rho}(t) ] \; \hat{\rho}(t).
\end{align}
This is the same form of the Gompertz rule \eqref{Grule}. In this way, Bloch equation can be considered as a sort of Gompertz rule.

Next, let us consider the $\kappa$-deformed density operator:

\begin{align}
  \hat{\rho}_{\kappa}( \beta) :=  \sum_i  \kexp( - \beta  E_i) \vert \psi_i \rangle \langle \psi_i \vert =  \kexp( - \beta  \hat{H}).
\end{align}
This leads to the following $\kappa$-deformation of Bloch equation:

\begin{align}
  -\frac{\partial}{\partial \beta} \hat{\rho}_{\kappa}(\beta) =\sum_i  E_i \frac{\kexp( - \beta  E_i)}{\ku \left[(\kexp(-\beta E_i) \right]} \,
 \vert \psi_i \rangle \langle \psi_i \vert   
 = \frac{\hat{H}}{ \ku \left[ \kexp( -\beta \hat{H} )  \right] } \, \hat{\rho}_{\kappa}(\beta).
 \label{k-BlochEq}
\end{align}
Again by changing the parameter $\beta $ to $t = - \ln \beta$ and using the relation \eqref{der_kexp}, we have

\begin{align}
  \frac{d }{dt} \hat{\rho}_{\kappa}(t) = -\frac{\kln[ \hat{\rho}_{\kappa}(t) ]}{\ku[ \hat{\rho}_{\kappa}(t) ]} \; \hat{\rho}_{\kappa}(t),
\end{align}
which can be regarded as a $\kappa$-deformation of the Gompertz rule.

Differentiating \eqref{k-BlochEq} again with respect to $\beta$, we obtain the following nonlinear differential equation:

\begin{align}
  (1 + \kappa^2 \beta^2 \hat{H}^2) \frac{\partial^2  \hat{\rho}_{\kappa}(\beta) }{\partial \beta^2}+ \kappa^2 \beta \hat{H}^2  \frac{\partial  \hat{\rho}_{\kappa}( \beta) }{\partial \beta} - \hat{H}^2 \, \hat{ \rho}_{\kappa}(\beta) = 0.
\end{align}
This differential equation reminds us of the research work \cite{CRS11} of the quantum free particle on two-dimensional hyperbolic plane. The relevant two-dimensional Schr\"odinger equation is separable in
the $\kappa$-dependent coordinate system $(z_x, y)$ with $z_x := x / \sqrt{1+\kappa^2 y^2}$. The Schr\"odinger equation $\hat{H}_1 \Psi = e_1 \Psi$
for the first partial Hamiltonian $\hat{H}_1$ leads to the following differential equation with the variable $z_x$ alone:
\begin{align}
  (1 + \kappa^2 z_x^2 ) \frac{ d^2 \Psi(z_x)}{d z_x^2} + \kappa^2 z_x  \frac{ d \Psi(z_x)}{d z_x}+ \mu \Psi(z_x) = 0,
  \quad \mu := \frac{2 m}{\hbar^2} e_1.
  \label{kSeq}
\end{align}
In the limit of $\kappa \to 0$, this differential equation reduces to the standard time-independent Schr\"odinger equation:
$d^2 \Psi(x) / d x^2 + \mu \Psi(x) = 0$.
 Cari\~nena et al. \cite{CRS11} obtained the solution of the differential equation \eqref{kSeq} as the $\kappa$-deformed plane wave (in our notations):

\begin{align}
  \Psi(z_x) = \exp\left[ \pm i \; \frac{\mu}{\kappa} \arsinh( \kappa \, z_x) \right], 
\end{align}
which is regarded as an $\arsinh$-type deformation.





\section{The $\kappa$-addition and the law of large number}

Next, we consider the $\kappa$-addition from the viewpoint of the law of large numbers (LLN), which plays a central role in probability, statistics, and
statistical physics \cite{LPS95}.
The $\kappa$-addition \cite{K13} is defined by

\begin{align}
 x \ksum y := x \sqrt{1+\kappa^2 y^2} + y \sqrt{1+\kappa^2 x^2}.
 \label{ksum}
\end{align}
This deformation of additive rule comes from the addition rule of the inverse hyperbolic sine function
as follows. For $a, b \in \mathbb{R}$, the addition rule is written as

\begin{align}
  \arsinh(a) + \arsinh(b) = \arsinh \left( a \sqrt{1+b^2} + b \sqrt{1+a^2} \right).
\end{align}
By setting $a = \kappa x$ and $b= \kappa y$, we obtain

\begin{align}
  \arsinh(\kappa x) &+ \arsinh(\kappa y) 
   = \arsinh \left( \kappa x \sqrt{1+\kappa^2 y^2} + \kappa y \sqrt{1+\kappa^2 x^2} \right) \notag \\
   &= \arsinh\Big[ \kappa ( x \ksum y ) \Big].
     \label{add_arsinh}
\end{align}
This relation is equivalent to the definition \eqref{ksum}.
The additive relation \eqref{add_arsinh} is readily generalized to
\begin{align}
  \sum_{i=1}^n \arsinh(\kappa x_i) = \arsinh\Big[ \kappa ( x_1 \ksum x_2 \ksum \cdots \ksum x_n ) \Big].
\end{align}
By applying this relation to the Boltzmann factor $\exp \left[ -\beta \sum_{i=1}^n K_{\kappa \beta}(p_i) \right] $
with respect to the $\kappa$-deformed kinetic energy \cite{WSM20} with $m=1$:

\begin{align}
 \sum_{i=1}^n  K_{\kappa \beta}(p_i) :=  \sum_{i=1}^n  \frac{1}{\kappa \beta} \arsinh \left(\kappa  \beta \frac{p_i^2}{2} \right),
\end{align}
we have

\begin{align}
  \exp \left[ -\beta \sum_{i=1}^n K_{\kappa \beta}(p_i) \right] &=
    \exp \left[ -\frac{1}{\kappa} \arsinh \left\{ \kappa \left( \beta \frac{p_1^2}{2} \ksum \beta \frac{p_2^2}{2} \ksum \dots \ksum \beta \frac{p_n^2}{2}  \right) \right\} \right] \notag \\
   &= \kexp \left[  \left(-\beta \frac{p_1^2}{2} \right) \ksum \left(-\beta \frac{p_2^2}{2} \right) \ksum \dots \ksum \left(-\beta \frac{p_n^2}{2}  \right) \right] \notag \\
   &= \kexp \left[ -\beta \frac{p_1^2}{2} \right] \kexp \left[-\beta \frac{p_2^2}{2} \right] \dots \kexp \left[-\beta \frac{p_n^2}{2}  \right] 
   = \prod_{i=1}^{n} \kexp \left[ -\beta \frac{p_i^2}{2} \right].
\end{align}
Note that the $\kappa$-exponential of the $\kappa$-summation of each term $-\beta  \frac{p_i^2}{2}$ in the second line is 
expressed as a factorized form in the last line.

It is well known that  LLN plays a fundamental role in statistical physics  \cite{LPS95}.
\L api\'nski \cite{LLN} showed that the standard LLN yields the most probable state of the system, which equals to the point of maximum of the entropy
and this point can be either Maxwell-Boltzmann statistics or Bose-Einstein statistics, or Zipf-Mandelbort law. 
McKeague \cite{Mc15} studied the central limit theorems under the special theory of relativity based on the $\kappa$-additivity. Scarfone \cite{S17}
studied the $\kappa$-deformation of Fourier transform and discussed the limiting distribution of the $\kappa$-sum of statistically independent variables.
The $\kappa$-additivity extension of the strong LLN is shown in \cite{Mc15} and it states that if $X_i$ are iid with finite mean, then

\begin{align}
\frac{X_1}{n} \ksum \frac{X_2}{n} \ksum \dots \ksum \frac{X_n}{n} \quad \to \quad \frac{1}{\kappa} \arsinh \left[ \kappa \ave{X} \right] {}_{a.s.},
\label{kLLN}
\end{align}
where a.s. stands for almost surely, i.e., the above sequence of the random variables $X_i$ converges almost surely, and $\ave{X}$ is the standard average of the random variable $X$.
Of course, in the limit of $\kappa \to 0$, the relation \eqref{kLLN} reduces to the standard strong LLN.
Note that the converged value in \eqref{kLLN} is the $\arsinh$-type deformation of the average $\ave{X}$.
In this way, the $\kappa$-additivity extension of the strong LLN suports the $\arsinh$-type deformation
of the average of a stochastic variable $X$.

\section{Contact density dynamics}

Nos\'e-Hoover (NH) thermostat \cite{NH} is a famous deterministic algorithm for constant-temperature molecular dynamics simulations.
Based on the idea of NH thermostat, several improved versions are proposed. Among them, contact density dynamics (CDD) \cite{CDD} is an algorithm based on contact Hamiltonian systems and generates any prescribed target distribution in physical phase space. The dynamical equations of CDD are the following.

\begin{subequations}
\begin{empheq}[left=\empheqlbrace]{align}   \frac{d q^i}{dt} &= \frac{\partial h(q, p, S)}{\partial p_i}, \label{v2p}\\
   \frac{d p_i}{dt} &= -\frac{\partial h(q, p, S)}{\partial q^i} + \frac{\partial h(p, q, S)}{\partial S} p_i, \\
   \frac{d S}{dt} &=- p_i \frac{\partial h(q, p, S)}{\partial p_i} + h(q, p, S), 
\end{empheq}
\end{subequations}
where $S$ is  the thermostatting variable, $q_i$ and  $p_i$ are the $i$-th component ($i=1, 2, \cdots, n$) of $n$-dimensional vectors, respectively. Here $h(q, p, S)$ denotes the contact Hamiltonian which is formed as

\begin{align}
  h(q, p, S) = \left( \rho_t(q, p) f(S) \right)^{-\frac{1}{n+1}} ,
\end{align}
with a target distribution $\rho_t(q, p)$ on $2n$-dimensional $\Gamma$-space and a normalized distribution $f(S)$ for the thermostatting variable $S$.
As in the case of Ref. \cite{NH}, we also choose $f(S)$ as the logistic distribution with scale $1$ and mean $c=0.0$:

\begin{align}
   f(S) = \frac{ \exp(S-c)}{(1 + \exp(S-c))^2}.
\end{align}

Utilizing this CDD algorithm, the $\kappa$-deformed exponential distributions are simulated. The target distribution $\rho_t(q, p)$ is the one-dimensional ($n=1$)
$\kappa$-deformed Gaussian function:

\begin{align}
  \rho_t(q, p) &= \frac{1}{Z_{\kappa}(\beta)} \exp \left[ -\beta H_{\kappa}(q, p) \right]
  = \frac{1}{Z_{\kappa}(\beta)}  \exp \left[ -\frac{1}{\kappa} \arsinh \left(\kappa  \beta \frac{p^2}{2} \right) \right] \exp\left[  -\beta \frac{q^2}{2} \right],
  \label{target}
\end{align}
where the associated Hamiltonian is

\begin{align}
H_{\kappa}(q, p)  &= \frac{1}{\kappa \beta} \arsinh \left(\kappa  \beta \frac{p^2}{2} \right) + \frac{q^2}{2},
\label{Hk}
\end{align}
and the normalization factor $Z_{\kappa}(\beta)$ \cite{K13} is

\begin{align}
Z_{\kappa}(\beta) = \frac{\pi}{\beta}
\frac{\sqrt{\frac{2}{ \kappa}} \Gamma \left(\frac{1}{2 \kappa }-\frac{1}{4}\right)}{\left(\frac{\kappa }{2}+1\right) \Gamma \left(\frac{1}{4}+\frac{1}{2 \kappa }\right)}.
 \end{align}

In general, the kinetic energy can be defined by
\begin{align}
  K(p) := \int_0^p v(p) dp,
 \end{align}
 where $v(p)$ denotes the constitutive relation between the velocity $v$ and the canonical momentum $p$. In the standard case of $v(p) = p/m$ with $m=1$,
 we have $K(p) = p^2/ 2$.
In the case of the Hamiltonian \eqref{Hk}, from \eqref{v2p} we have

\begin{align}
  v_{\kappa}(p) := \frac{d q}{dt}= \frac{\partial H_{\kappa}(q, p)}{\partial p}  &= 
  \frac{ p }{ \ku \left[ \kexp \left(- \beta \frac{p^2}{2} \right) \right]} =  \frac{ p }{ \sqrt{1+  \kappa^2\left(\beta \frac{p^2}{2} \right)^2}}.
  \label{v(p)}
 \end{align}
 It is worthwhile to note that the $v_{\kappa}(p)$ has a $\beta$ (or temperature) dependency when $\kappa \ne 0$.
Then the corresponding kinetic energy $K_{\kappa}(p)$ is the first term $ \frac{1}{\kappa \beta} \arsinh \left(\kappa  \beta \frac{p^2}{2} \right)$
 in \eqref{Hk},
  which can be regarded as a $\kappa$-deformation of the standard kinetic energy $p^2/2$.

\begin{figure}[H]
\includegraphics[width=7 cm]{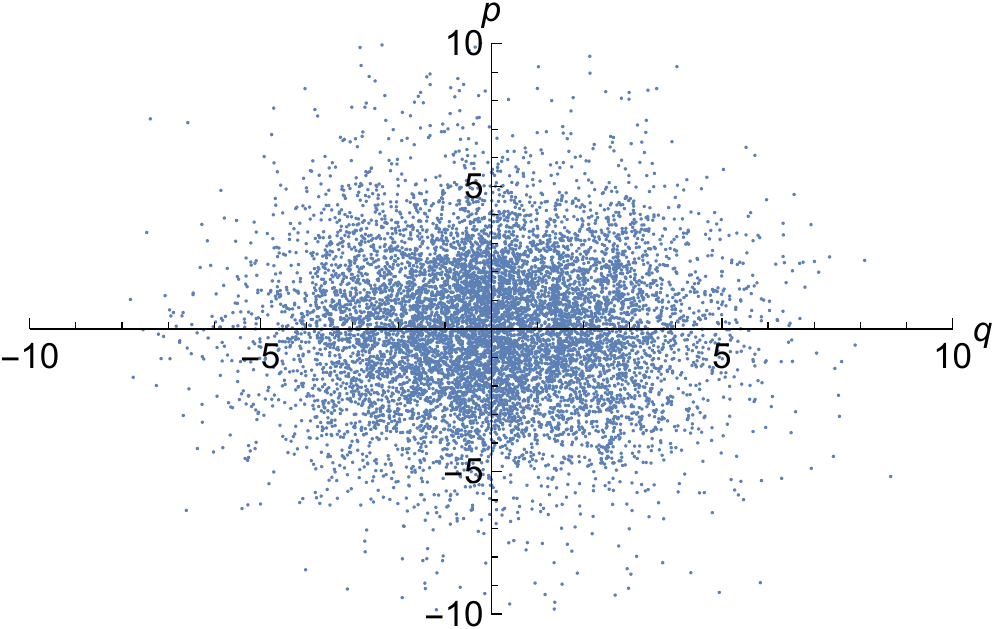}
\hspace{4mm}
\includegraphics[width=6 cm]{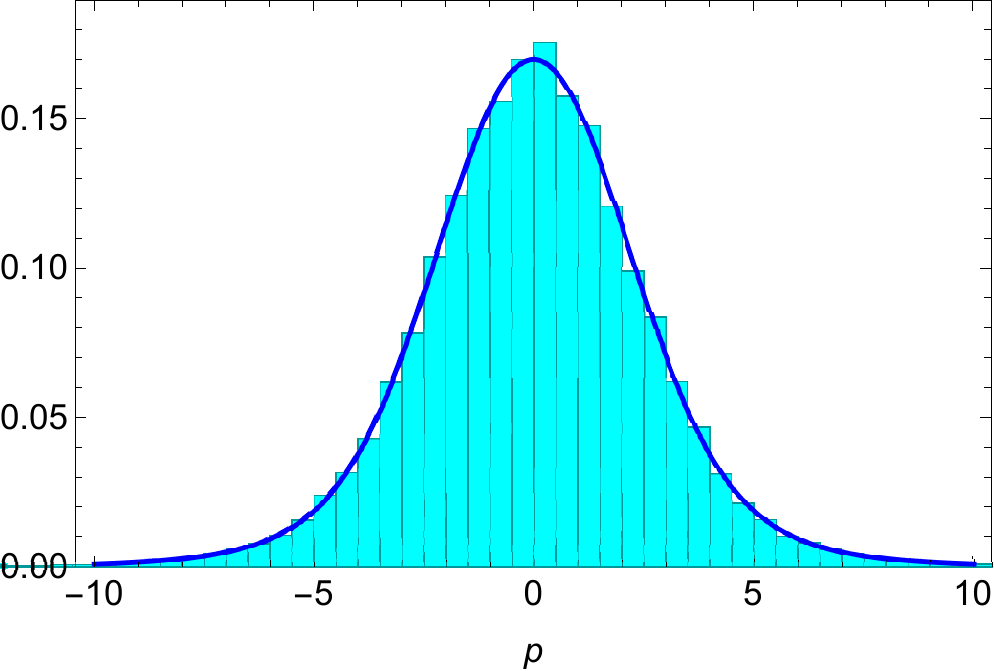}
\caption{The simulated results of the CDD simulations of the target distribution \eqref{target} with $\kappa=0.4$ and $\beta=0.2$. (\textbf{a}) the phase ($q$-$p$) space orbit of the $\kappa$-deformed distribution. The $1.5 \times10^4$ points of a simulated orbit with the initial condition ($q_0=0.1, p_0=0.1$, and $S_0=0.9$ are shown. (\textbf{b}) the histogram ({\color{cyan} cyan bars}) of the frequencies for $p$ and the corresponding momentum $\kappa$-distribution ( {\color{blue} blue solid curve} ). 
}
\label{fig1}
\end{figure}

We have performed a number of the CDD simulations for the target state \eqref{target} with different parameters and initial conditions.
As an example, Figure~\ref{fig1} shows the phase space orbit and the histogram of the frequencies of the momentum $p$ for a typical result of the CDD
simulation of the target state \eqref{target} with $\beta=0.2$, $\kappa=0.4$. The initial conditions used are also denoted in the figure captions.

The CDD simulated result obeys the ergodicity as can be seen from the well distributed points in the phase space in Figure~\ref{fig1} (a). 
Note that the momentum distribution in the histogram of Figure~\ref{fig1} (b) is well fitted with the $\kappa$-Gaussian distribution,
which is cased by the $\arsinh$-type deformation of the kinetic energy $p^2/2$.

Note also that for the $\kappa$-deformed Hamiltonian \eqref{target}, we have \cite{WSM20}

\begin{align}
   \ave{ p \frac{\partial }{\partial p} H_{\kappa}(q, p)} = \frac{1}{\beta}  
\label{this}
\end{align}
which reminds us of \textit{a generalization of equipartition theorem} \cite{To}: $\ave{ p \frac{\partial}{\partial p} \mathcal{H}} = k_{\rm B} T$,
where $\mathcal{H}$ is the Hamiltonian of a system in thermal equilibrium
with the temperature $T$.

\section{Conclusions}
We have considered the $\kappa$-deformations of some quantities concerning statistical physics and pointed out some unexpected relations
among different fields such as statistical mechanics, mathematical biology and evolutional game theory.
Especially, we focus on the $\arsinh$-type deformation of the ratio $\beta p^2/2$ of kinetic energy to the average thermal energy
$k_{\rm B} T = 1/\beta$.
With the help of the thermostat (CDD) algorithm, we have performed the relevant numerical simulations for the Hamiltonian with the $\arsinh$-type deformation of kinetic energy term and show the resultant momentum distribution is the $\kappa$-Gaussian distribution.

Finally, we would like to point out the relation which might be suggestive for future research.
Let us consider the $\kappa$-deformed energy density of state  $\Omega_{\kappa}(U)$:
\begin{align}
 \Omega_{\kappa}(U) := \kexp \left( \frac{U}{k_{\rm B} T_c} \right) = \exp \left[ \frac{1}{\kappa} \arsinh \left( \kappa \frac{U}{k_{\rm B} T_c} \right) \right],
\end{align}
which is the $\kappa$-deformation of the energy density of state $\exp(U/ k_{\rm B} T_c)$ for the thermal reservoir with a constant-temperature $T_c$ (Boltzmann reservoir \cite{BR}).
In other words, $\ln  \Omega_{\kappa}(U) $ is regarded as the $\arsinh$-type deformation of the ratio $U / (k_{\rm B} T_c)$.
The Boltzmann temperature $T(U)$ for this $\kappa$-deformed thermal reservoir is given by

\begin{align}
 \frac{1}{k_{\rm B} T(U) } :=  \frac{d \ln \Omega(U)}{d U} 
 = \frac{ \frac{1}{k_{\rm B} T_c} }{ \sqrt{ 1 +  \kappa^2 \left( \frac{U}{k_{\rm B} T_c} \right)^2}}.
\end{align}
Rearranging this relation leads to

\begin{align}
 k_{\rm B} T(U)  &=   \sqrt{ (\kappa  U)^2 +  (k_{\rm B} T_c )^2}, 
\end{align}
which reminds us of the relativistic energy-momentum relation: $E(p) = \sqrt{ (c p)^2 +  (m c^2)^2}$.

%

\vspace{6pt} 



\authorcontributions{
Conceptualization, T.W. and A.S.; methodology, T.W.; software, T.W.; validation, T.W. and A.S.; formal analysis, T.W.; investigation, T.W. and A.S.; resources, T.W.; data curation, T.W.; writing---original draft preparation, T.W.; writing---review and editing, T.W. and A.S.; visualization, T.W.; supervision, T.W.; project administration, T.W.; funding acquisition, T.W. All authors have read and agreed to the published version of the manuscript.}

\funding{
The first named author (T.W.) is partially supported by Japan Society for the Promotion
of Science (JSPS) 
Grants-in-Aid for Scientific Research (KAKENHI)
Grant Number 22K03431. 
}


\informedconsent{Not applicable.}

\dataavailability{
Not applicable.
} 

\acknowledgments{The authors thank anonymous referees for their valuable comments.}

\conflictsofinterest{The authors declare no conflict of interest.
} 





\appendixstart
\appendix
\section[Appendix A]{Basics of the $\kappa$-deformed functions}
\label{Appendix A}

Here we briefly review some $\kappa$-deformed functions and the associated useful relations \cite{K02, K05}.
Because all $\kappa$-deformed functions are symmetric under the sign change of the deformation parameter $\kappa$, i.e.,  changing $\kappa$ to $-\kappa$, throughout this paper we assume $\kappa > 0$.
In the $\kappa \to 0$ limit, the $\kappa$-exponential function \eqref{k-exp} and the $\kappa$-logarithmic function \eqref{k-ln} reduce to the standard exponential function $\exp(x)$ and
logarithmic function $\ln (x)$, respectively.

\begin{align}
 \lim_{\kappa \to 0} \kexp(x) = \exp(x), \quad
 \lim_{\kappa \to 0} \kln x = \ln x.
\end{align}
We next introduce another $\kappa$-deformed function:

\begin{align}
 \ku(x) \equiv \frac{x^{\kappa} + x^{-\kappa}}{2}
 = \cosh \big[ \kappa \ln(x) \big],
\end{align}
which is the conjugate (or co-function) of $\kln x$, as similar
as that $\cos (x)$ is the co-function of $\sin (x)$. 
In the $\kappa \to 0$ limit, this $\kappa$-deformed
function reduces to the unit constant function $u_0(x)=1$.
By using $\ku(x)$,
the derivative of the $\kappa$-exponential is expressed as

\begin{align}
  \frac{d}{dx} \kexp(x)  = \frac{ \kexp(x)}{ \ku \left[ \kexp(x) \right]} = \frac{ \kexp(x)}{\sqrt{1 + \kappa^2 x^2}} ,
\label{der_kexp}
\end{align}
and the derivative of $\kappa$-logarithm is expressed as

\begin{align}
  \frac{d}{dx} \kln(x)  = \frac{ \ku(x)}{x},
\end{align}
respectively.

The inverse function of $\ku(x)$ is 

\begin{align}
 \ku^{-1}(x) = \exp \left[ \arcosh (\kappa x) \right],
\end{align}
that is the co-function  of $\kexp(x)$.

The $\kappa$-entropy $\Sk$ \cite{K02, K05} is a $\kappa$-generalization of the Gibbs-Shannon entropy $S^{\rm GS}=-k_{\rm B} \sum_i p_i\,\ln p_i$ by replacing
the standard logarithm with the $\kappa$-logarithm, i.e.,

\begin{align}
  \Sk =-k_{\rm B} \sum_i p_i\, \kln p_i.
\end{align}



\section[Appendix B]{Replicator equations and the general form of Lotka-Volterra equations}
\label{Appendix B}

We here summarize some known important facts in mathematical biology and evolutional game theory according to Ref. \cite{Harper09,Harper11,Baez}.
Consider a discrete probability distribution described by a set of $n$ positive variables $\boldsymbol{x}=(x_1, x_2, \ldots, x_n)$ with the normalization
$\sum_i^n x_i = 1$, where each $x_i$ denotes the proportion of the $i$-th type in the total population.
The RE for this distribution is given by

\begin{align}
  \frac{d}{dt} x_i = x_i \Big( f_i(\boldsymbol{x}) - \bar{f}(\boldsymbol{x}) \Big),
  \label{RE}
\end{align}
where $ f(\boldsymbol{x}) = (f_1(\boldsymbol{x}), \ldots, f_n(\boldsymbol{x}))$ is a fitness landscape and $\bar{f}(\boldsymbol{x}) = \sum_{i=1}^n x_i f_i(\boldsymbol{x})$ is the mean fitness.
Replicator dynamics can be described as a time evolutional curve
on the simplex $\Delta^n := \{ \boldsymbol{ x} \in {\mathbb R}^n_+  \; \vert \;  x_i \ge 0, \sum_i x_i = 1 \}$ with the matrix component $ g^{ij} ( \boldsymbol{x})$ of \textit{Shahshahani metric} \cite{Sigmund87} $g$ as

\begin{align}
    g^{ij} ( \boldsymbol{x}) = \frac{\delta_{ij}}{x_i},  
\end{align}
The inverse matrix is $g_{ij}(\boldsymbol{x}) = x_i \delta_{ij}$.
Note that the $n$-simplex $\Delta^n$ is $(n-1)$-dimensional and the
Shahshahani metric diverges on the boundary of the simplex.
So this metric is valid only on the interior $S^n$ of $\Delta^n$.

There is a natural mapping: $(p_1, p_2, \ldots, p_n) \to (x_1, x_2, \ldots, x_n)$. Fisher metric is induced by the Shahshahani metric under this mapping.
\begin{align}
  (g^{\rm F})^{ij}(\boldsymbol{x}) = \Ev{}{\frac{\partial \ln \boldsymbol{x}}{\partial x_i}
    \frac{\partial \ln \boldsymbol{x}}{\partial x_j}}
  = \sum_{k=1}^n x_k \frac{\delta_{ik}}{x_i} \frac{\delta_{ik}}{x_i}
  = \frac{\delta_{ij}}{x_i}.
\end{align}

It is known that the Shahshahani manifolds yields an interpretation of the RE.
Theorem 1 in \cite{Harper09}: if the differential equation $d x_i /dt = f_i(\boldsymbol{x})$ is a Euclidean gradient with $f_i = \partial V / \partial x_i$, the RE \eqref{RE}
is a gradient with respect to Shahshahani metric.
A brief explanation is as follows.
The gradient with respect to Shahshahani metric is

\begin{align}
  (\nabla_g V)_i = \sum_j g_{ij} \frac{\partial V}{\partial x_j}
  = \sum_j x_i \delta_{ij} f_j = x_i f_i, 
\end{align}
which is the first term in the left hand side of the RE \eqref{RE}.
The variable $x_i$ in the RE has to satisfy
the normalization constraint ($\sum_i x_i = 1$), i.e.,
the dynamics of each $x_i$ is restricted on the simplex $\Delta^n$.
Recall that Shahshahani metric is valid only on the interior $S^n$ of $\Delta^n$.
Indeed, the normalization constraint is satisfied during an time evolution
as follows

\begin{align}
  \frac{d}{dt} \sum_i x_i = \sum_i \frac{d x_i}{dt} = \sum_i x_i ( f_i - \bar{f} )
  = \sum_i x_i f_i - \bar{f} = 0.
\end{align}

The state $\hat{\boldsymbol{x}}$ is said to be \textit{evolutionarily stable state} if for all $\boldsymbol{x} \ne \hat{\boldsymbol{x}}$ in some neighborhood of $\hat{\boldsymbol{x}}$,

\begin{align}
   \boldsymbol{x \cdot f}(\boldsymbol{x}) < \hat{\boldsymbol{x}} \boldsymbol{\cdot f}(\boldsymbol{x}).
\end{align}
Let the potential $V(\boldsymbol{x}) = \mathrm{D}(\hat{\boldsymbol{x}} \Vert \boldsymbol{x})
= \sum_i \hat{x}_i \ln \hat{x}_i - \sum_i \hat{x}_i \ln x_i$, then we have

\begin{align}
  \frac{d}{dt} V(\boldsymbol{x}) = - \sum_i \hat{x}_i \frac{1}{x_i} \frac{d x_i}{dt}
  = -\sum_i \hat{x}_i ( f_i - \bar{f}) = -\sum_i \hat{x}_i f_i + \bar{f}
  = - (\hat{\boldsymbol{x}} \boldsymbol{\cdot f} - \boldsymbol{x \cdot f}) < 0.
\end{align}
Hence the Kullback-Leibler divergence $\mathrm{D}(\hat{\boldsymbol{x}} \Vert \boldsymbol{x})$ is a local Lyapunov function for the RE.


Next, if $x_i = \exp( v_i(\boldsymbol{x}) - \psi )$ with $d v_i(\boldsymbol{x}) /dt = f_i(\boldsymbol{x})$ and $\psi(\boldsymbol{x})$ a normalization constant. From the normalization $\sum_i x_i =1$, we have

\begin{align}
  0 = \sum_i \frac{d}{dt} x_i = \sum_i \left( \frac{d}{dt} v_i(\boldsymbol{x}) - \frac{d}{dt} \psi(\boldsymbol{x}) \right) x_i = \sum_i x_i f_i(\boldsymbol{x}) - \frac{d}{dt} \psi(\boldsymbol{x}) = \bar{f}(\boldsymbol{x}) - \frac{d}{dt} \psi(\boldsymbol{x}).
\end{align}
As a result we see that $d \psi(\boldsymbol{x}) / dt = \bar{f}(\boldsymbol{x})$, and $x_i$ satisfies

\begin{align}
  \frac{d}{dt} x_i = x_i \left( \frac{d}{dt} v_i(\boldsymbol{x}) - \frac{d}{dt} \psi(\boldsymbol{x}) \right)
  = x_i ( f_i(\boldsymbol{x}) - \bar{f}(\boldsymbol{x}) ).
\end{align}
Consequently, the exponential families $x_i = \exp( v_i(\boldsymbol{x}) - \psi )$ are solutions of the RE.

If there is no constraint the corresponding dynamics is described by the gLV equation \eqref{gLVeq}.
The gLV equations and REs are related as follows. Let each $y_i$ satisfies the gLV equation \eqref{gLVeq}. 
Changing the variable $y_i$ to $x_i$ as

\begin{align}
   x_i = \frac{y_i}{\sum_{j=1}^n y_j},
   \label{x_i}
\end{align}
which lead to the new normalized variables $\{ x_i \}$, i.e., $\sum_j x_j = 1$. Then, we see that

\begin{align}
   \frac{ d x_i}{dt} = \frac{\frac{d y_i}{dt}}{\sum_j y_j} - y_i \frac{\sum_k \frac{d y_k}{dt}}{\left( \sum_j y_j \right)^2}
  = \frac{y_i f_i}{\sum_j y_j} - \frac{y_i}{\left( \sum_j y_j \right)} 
  \frac{\sum_k y_k f_k}{\left( \sum_j y_j \right)}
 = x_i ( f_i - \bar{f} ).
\end{align}
Thus, the transformed variable $x_i$ in \eqref{x_i} satisfies the RE.

\begin{adjustwidth}{-\extralength}{0cm}

\reftitle{References}

\end{adjustwidth}
\end{document}